# Noncommuting Observables and Local Realism

James D. Malley[1] and Arthur Fine[2]

[1]*Center for Information Technology, National Institutes of Health,*
[2]*University of Washington, Seattle*

A standard approach in the foundations of quantum mechanics studies local realism and hidden variables models exclusively in terms of violations of Bell-like inequalities. Thus quantum nonlocality is tied to the celebrated no-go theorems, and these comprise a long list that includes the Kochen-Specker and Bell theorems, as well as elegant refinements by Mermin, Peres, Hardy, GHZ, and many others. Typically entanglement or carefully prepared multipartite systems have been considered essential for violations of local realism and for understanding quantum nonlocality. Here we show, to the contrary, that sharp violations of local realism arise almost everywhere without entanglement. The pivotal fact driving these violations is just the noncommutativity of quantum observables. We demonstrate how violations of local realism occur for arbitrary noncommuting projectors, and for arbitrary quantum pure states. Finally, we point to elementary tests for local realism, using single particles and without reference to entanglement, thus avoiding experimental loopholes and efficiency issues that continue to bedevil the Bell inequality related tests.

**I. INTRODUCTION** Bell-style arguments involving entanglement and carefully coupled systems have together been at first catalyzing agents and then, more recently, stunning resources for quantum information theory; see [1, 2]. Typically, experimental and theory-based study of quantum nonlocality and violations of local realism are framed in these terms. It turns out, however, that neither entanglement nor the violation of the Bell inequalities are essential for these no-go results. Rather, as we will show, the framework within which local realism is defined (and Bell-like results derived) must already fail given a single pair of noncommuting observables, regardless of whether the system is in an entangled state for which the Bell inequalities are violated, or is even a composite system at all. It follows from the results below that the framework of local realism fails even for certain product composite states $D = D_1 \otimes D_2$ where there is no entanglement and no violation of the Bell inequalities. It also fails, as we shall show, for a single qubit. Thus the fate of local realism, along with that of noncontextual hidden variables, is sealed by noncommutativity alone. If the concern is to show the difficulty with these programs, strong no-go theorems require neither entanglement, nor careful pair production with spacelike separated measurements, nor highly efficient detection.

In more detail, no-go theorems constrain the assignment of premeasurement values so as to insure either noncontextuality or, in the case of coupled systems, locality. In the style of the Kochen-Specker (KS) theorems (noncontextual hidden variables) the constraints are algebraic (sum, product rules, etc.) In the style of the Bell theorems (local realism) the constraints impose independence conditions ("shared randomness") for interacting systems in particular composite states. Whether in the KS style or the Bell style the constraints that enable the no-go theorems are equivalent to a certain uniform way of obtaining the quantum statistics in a given state [1, 2].

With respect to a given state (density operator) $D$, both KS and Bell require that the relevant quantum observables can be represented as random variables, all defined on a common space, whose single and joint distributions agree with the quantum single probability distributions in $D$ and with the quantum joint distributions for commuting pairs in $D$. We will call this requirement *BKS(D)* (or *BKS($\varphi$)*, confounding, as usual, pure states $\varphi$ and density operators $|\varphi\rangle\langle\varphi|$) and, importantly, we take the relevant observables to include at least all those represented by one-dimensional projectors.

Our aim is to fully identify the logical engine driving the no-go theorems with the most basic, nonclassical feature of the quantum theory: that not all observables commute. The program of trying to make sense of nonlocal or contextual phenomena for systems where the standard Bell (or KS) conditions fail is, therefore, exactly the ultimately unsatisfying program of trying to denature noncommuting objects into commuting ones.

That noncommutativity may be the culprit was suggested by a recent result showing that if $BKS(D)$ were to hold globally (i.e., for *every* state $D$) then *all* observables would commute [3]. Here, we develop a technique involving the compound commutator (itself an observable) that requires much weaker, perhaps minimal assumptions. Using that technique we are able to get beneath this global result to pinpoint the conflict between noncommutativity and the *BKS* conditions to each state individually and to each single pair of noncommuting projectors.

**II. *BKS* AND COMMUTATIVITY.** We do this starting from either of two sets of initial data: *(I)* we are given any two noncommuting projectors $A$, $B$ [and then construct a certain state $\varphi$ that depends on $A$, $B$]; or, *(II)* we are given an arbitrary state $\varphi$ [and then define a specific pair of noncommuting projectors $A$, $B$ that depend on $\varphi$]. In either situation *(I)* or *(II)* it follows that $BKS(\varphi)$ entails that the projectors in question misbehave with respect to commutativity. We begin by recalling results from [3, 6].

*Lemma 1*. If $BKS(\varphi)$ holds and neither $tr[DA]$ nor $tr[DB]$ is zero, then the relation

$$tr[DABA] = tr[DBAB] \tag{2.1}$$

is valid for the system in state $D = P_\varphi = |\varphi\rangle\langle\varphi|$.

This is just a restatement of *Theorem 1* in [3]. The derivation of (2.1) uses a simple version of Gleason's theorem (due to Gudder) for one-dimensional projectors; see [4; Corollary 5.17]. This version and the proof of (2.1) involve only elementary facts about additive functions and inner product spaces.

Given projectors $A$ and $B$, we make use of a relation between the usual commutator $C = [A, B] = AB - BA$, and the compound commutator $[AB, BA]$, specifically that

$$C^*C = [A,B]^*[A,B] = (A - B)[AB, BA], \tag{2.2}$$

valid since $A^2 = A$, $B^2 = B$, as both are projectors. It is worth noting that

$$G = G(A,B) \equiv [AB, BA] = ABA - BAB, \tag{2.3}$$

is an observable, while $[A, B]$ is not. We now obtain

*Lemma 2.* If (2.1) holds for $D = |\varphi\rangle\langle\varphi|$, where $G(A,B)\varphi = \lambda\varphi$, for some real $\lambda$, then $[A, B]\varphi = 0$.

*Proof.* From (2.1) we get $0 = tr[D(ABA - BAB)] = tr[DG] = \langle\varphi|G|\varphi\rangle = 0$. But if $G\varphi = \lambda\varphi$ for any real $\lambda$, then $\langle\varphi|G|\varphi\rangle = \lambda$, and so $G\varphi = \lambda\varphi = 0$. From (2.2) with $C = [A,B]$, we have $C^*C\varphi = 0$, so $\langle\varphi|C^*C|\varphi\rangle = \langle C\varphi|C\varphi\rangle = 0$. Hence $C\varphi = [A,B]\varphi = 0$. ∎

Next, let $S(A, B)$ be a full set of distinct eigenvectors, $\varphi$, associated with the nonzero eigenvalues appearing in the spectral decomposition of the observable $G(A, B)$. Then our first main result is the following, where we assume the initial data of situation *(I)*:

*Theorem 1.* Suppose given noncommuting projectors $A, B,$ in a quantum system. Then

(1) $S(A, B)$ is not empty; (2) for any state $\varphi$ in $S(A, B)$, we have $(AB - BA)\varphi \neq 0$;

(3) for any state $\varphi$ in $S(A, B)$, $BKS(\varphi)$ contradicts (2) and so cannot be valid.

*Proof.* We first show (1), that $S(A, B)$ is not empty. If it were then $G = G(A,B)$ would be identically zero. But then from (2.2) and (2.3) it would follow that $[A,B]^*[A,B] = 0$, and hence that $[A, B] = 0$; i.e., that $A$ and $B$ commute.

Now, to prove (2) let $\varphi \in S(A, B)$, so that $G\varphi = \lambda\varphi$, for real eigenvalue $\lambda \neq 0$. Assume to the contrary, that $C\varphi = (AB - BA)\varphi = 0$. Using (2.2), (2.3) and the fact that $\varphi \in S(A, B)$, we see that $0 = C^*C\varphi = (A - B)G\varphi = (A - B)(\lambda\varphi) = \lambda(A - B)\varphi$, for $\lambda \neq 0$. Thus $A\varphi = B\varphi = \eta$, say. Then also $A\eta = A(A\varphi) = A^2\varphi = A\varphi = \eta$; similarly, $B\varphi = B\eta = \eta$. Hence $G\varphi = (ABA - BAB)\varphi = ABA\varphi - BAB\varphi = AB\eta - BA\eta = A\eta - B\eta = \eta - \eta = 0$. However $G\varphi = \lambda\varphi \neq 0$. Hence only $C\varphi \neq 0$ is possible, which verifies (2).

For the proof of (3) assume again that $\varphi \in S(A, B)$. By considering cases we show that necessarily $A\varphi \neq 0$, $B\varphi \neq 0$, irrespective of the status of $BKS(\varphi)$. Thus, *Case (a)*: if $A\varphi = 0, B\varphi = 0$, we would have $\lambda\varphi = G\varphi = (ABA - BAB)\varphi = 0$, while $\lambda \neq 0$, and this is a contradiction; *Case (b)*: if $A\varphi = 0, B\varphi \neq 0$, then $G\varphi = (ABA - BAB)\varphi = BAB\varphi = \lambda\varphi$, so we would have $(BAB)(BAB)\varphi = \lambda^2\varphi \neq 0$. But $(BAB)(BAB) = BABAB$, and $BABAB\varphi = BA(BAB\varphi) = BA(\lambda\varphi) = \lambda(BA\varphi) = 0$, which is again a contradiction. A similar problem arises in the remaining case where $A\varphi \neq 0, B\varphi = 0$, (look at $(ABA)(ABA)$). We can assume, therefore, that $A\varphi \neq 0, B\varphi \neq 0$.

To finish the proof of (3) suppose now that $BKS(\varphi)$ holds, for $\varphi \in S(A, B)$, and that $A\varphi \neq 0, B\varphi \neq 0$. We can apply *Lemma 1* to see that (2.1) holds for the state $D = |\varphi\rangle\langle\varphi|$. But since $\varphi \in S(A, B)$, *Lemma 2* applies and $(AB - BA)\varphi = 0$. However from (2) we must also have $(AB - BA)\varphi \neq 0$. We conclude that $BKS(\varphi)$ cannot be valid. ∎

Let's now follow the consequences of starting with the initial data in situation *(II)*:

*Theorem 2*. To every pure state $\varphi$ there corresponds a projector pair $A, B$ such that (1) $\langle\varphi|G|\varphi\rangle \neq 0$; (2) $[A, B]\varphi \neq 0$; (3) $BKS(\varphi)$ contradicts (1) and (2) and so cannot be valid.

*Proof.* Let $A = P_\varphi = |\varphi\rangle\langle\varphi|,$ with $\langle\varphi|\varphi\rangle = 1$. Pick any $\xi$ orthogonal to $\varphi$, and let $B = (1/2)|\varphi + \xi\rangle\langle\varphi + \xi|.$ Since $\langle\varphi + \xi|\varphi + \xi\rangle = 2,$ $B$ is a projector. Then, direct calculation shows that $A\varphi = \varphi,$ $B\varphi = (1/2)(\varphi + \xi),$ $[A, B]\varphi = -(1/2)\xi \neq 0,$ and $\langle\varphi|G|\varphi\rangle = 1/4 \neq 0.$ This verifies (1) and (2).

For (3) note that by construction $A\varphi \neq 0, B\varphi \neq 0,$ so if $BKS(\varphi)$ were to hold then *Lemma 1*, and therefore (2.1) would apply. But then for $D = |\varphi\rangle\langle\varphi|$ we would have $0 = tr[DABA] - tr[DBAB] = tr[D(ABA - BAB)] = \langle\varphi|G|\varphi\rangle.$ However, this means $\langle\varphi|G|\varphi\rangle = 0,$ which contradicts (1).

On the other hand, $A\varphi = \varphi$ and $B^2 = B$ implies $BG\varphi = B(ABA - BAB)\varphi = (BAB)(A\varphi) - BAB\varphi = BAB\varphi - BAB\varphi = 0,$ so that $\langle\varphi|BG|\varphi\rangle = 0.$ Moreover, $A\varphi = \varphi$ also means $\langle\varphi|AG|\varphi\rangle = \langle\varphi|G|\varphi\rangle.$ If now $BKS(\varphi)$ is assumed to hold we must have $\langle\varphi|G|\varphi\rangle = 0,$ so $\langle\varphi|AG|\varphi\rangle = 0.$ Hence $\langle\varphi|C^*C|\varphi\rangle = \langle\varphi|(A - B)G|\varphi\rangle = \langle\varphi|AG|\varphi\rangle - \langle\varphi|BG|\varphi\rangle = 0.$ Thus, under $BKS(\varphi)$ we get $C\varphi = [AB - BA]\varphi = 0,$ which contradicts (2). ∎

**III. EXTENSION TO QUBITS.** Recall that Gleason provides a density-trace relation for probability measures on the space of projectors, by which $\Pr_D(A) = tr[DA]$ when the system is in state $D$. Busch [5] has shown that if the usual projector valued measure (*PV*) is extended to include a positive operator valued measure (*POVM*), then the existence of the density $D$ as in Gleason's result is again assured even for $dim = 2$. Thus suppose we take as relevant observables for $BKS$ all positive, semidefinite operators (which includes all the projectors) and call the resulting assumption of a random variables representation

*BBKS*. Then in *Lemma 1* if we replace *BKS*($\varphi$) with *BBKS*($\varphi$), everything goes through as before. The same is true for *Lemma 2* and *Theorems 1* and *2*. Thus even for a single qubit, noncommutativity stands in the way of local, or noncontextual hidden variables provided we allow the full set of *POVMs* to count as the class of measurements. Note that the toy hidden variables models of Bell and of KS for *dim* = 2 assume *BKS* applies only for the space of orthogonal projectors.

**IV. EXPERIMENTAL IMPLICATIONS.** We confine the discussion to tests for locality. These involve coupled systems where the constraints defining locality imply Bell inequalities. Where these are violated experimentally, nonlocality is implicated. In this context, however, these same locality constraints imply the satisfaction of the *BKS* conditions. As we've shown, these, in turn, imply that certain noncommuting observables (depending on the state of the coupled system) commute (on that state). Hence elementary tests for commutativity can show that the *BKS* conditions fail, so that locality is violated. These tests are not only simpler than tests for the Bell inequalities but, because they apply to coupled systems even in states where the Bell inequalities (or related conditions) may hold, they open up a wider range of cases where it can be shown experimentally that local realism must fail. Several such tests are described in [6], one being related to the classical Young double slit experiment where interference is a consequence of the noncommutativity, while another involves sequential tests on single particles that probe for a proper joint distribution consistent with certain observed conditional distributions. Neither of these experiments require entanglement.

**V. DISCUSSION.** The existence of elementary experimental procedures for ruling out local realism, regardless of the composite state, follows from our two theorems that show how the *BKS* conditions, which for coupled systems are equivalent to the usual locality constraints, conflict with the facts regarding arbitrary noncommuting projectors, and for arbitrary quantum states. These striking results, however, depend on the general assumption of *BKS* (and so of local realism) at least for all one-dimensional projectors. That is a perfectly reasonable assumption if one is to take local realism seriously. The Bell inequalities, however, require less; namely, only that local realism (and so *BKS*) hold for the finite number of observables in a given experimental protocol (a minimum of four: two non-commuting pairs for each system). It may well be possible to refine our results even further so as to reduce the reliance on so many projectors, since it does appear that simple noncommutativity alone is local realism's poison pill. For the present however, one has a choice. Assuming *BKS* for lots of observables makes for widespread conflict with local realism (or noncontextual hidden variables), a conflict that can be easily tested experimentally: any pair of non-commuting observables and any state will show the conflict. Assuming *BKS* only for a handful of observables makes for a restricted conflict with local realism—we need Bell-violating coupled systems—but also demands next-generation experiments free of the efficiency or signaling loopholes of the current ones.

Finally, the sharp conflict with noncommutativity presented here makes essential use of the inner product machinery of Hilbert space and does not apply to models for quantum theory that do not, with the Bohm model being a primary case in point.